\documentclass[fleqn,usenatbib]{mnras}

\usepackage{newtxtext,newtxmath}
\usepackage{deluxetable}
\usepackage{natbib,amsmath,multirow,tabularx}

\usepackage[T1]{fontenc}

\DeclareRobustCommand{\VAN}[3]{#2}
\let\VANthebibliography\thebibliography
\def\thebibliography{\DeclareRobustCommand{\VAN}[3]{##3}\VANthebibliography}

\usepackage{graphicx}	
\usepackage{amsmath}	

\title[Revised M-dwarf Outlier Temperatures]{Revised Temperatures For Two Benchmark M-dwarfs -- Outliers No More}

\author[Martin et al.]{%
        David V. Martin$^{1,2*}$$^{\href{https://orcid.org/0000-0002-7595-6360}{\includegraphics[scale=0.5]{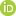}}}$,
        Tayt Armitage$^{1}$,
        Alison Duck$^{1}$$^{\href{https://orcid.org/0000-0002-4531-6899}{\includegraphics[scale=0.5]{orcid.jpg}}}$,
        Matthew I. Swayne$^{3}$$^{\href{https://orcid.org/0000-0002-2609-3159}{\includegraphics[scale=0.5]{orcid.jpg}}}$,
        Romy Rodr\'iguez Mart\'inez$^{1}$$^{\href{https://orcid.org/0000-0003-1445-9923}{\includegraphics[scale=0.5]{orcid.jpg}}}$, \newauthor
        Ritika Sethi$^{4}$,
        B. Scott Gaudi$^{1}$$^{\href{https://orcid.org/0000-0003-0395-9869}{\includegraphics[scale=0.5]{orcid.jpg}}}$,
        Sam Gill$^{5}$$^{\href{https://orcid.org/0000-0002-4259-0155}{\includegraphics[scale=0.5]{orcid.jpg}}}$,
        Daniel Sebastian$^{6}$$^{\href{https://orcid.org/0000-0002-2214-9258}{\includegraphics[scale=0.5]{orcid.jpg}}}$,
        Pierre F. L. Maxted$^{3}$$^{\href{https://orcid.org/0000-0003-3794-1317}{\includegraphics[scale=0.5]{orcid.jpg}}}$\newauthor
\\
*martin.4096@osu.edu\\
$^{1}$Department of Astronomy, The Ohio State University, 4055 McPherson Laboratory, Columbus, OH 43210, United States of America\\
$^{2}$NASA Sagan Fellow\\
$^{3}$Astrophysics Group, Keele University, Staffordshire, ST5 5BG, United Kingdom \\
$^{4}$Department of Physical Sciences, Indian Institute of Science Education and Research, Berhampur, Odisha 760010, India \\
$^{5}$Department of Physics, University of Warwick, Gibbet Hill Road, Coventry CV4 7AL, United Kingdom \\
$^{6}$School of Physics and Astronomy, University of Birmingham, Edgbaston, Birmingham B15 2TT, United Kingdom\\
\vspace{-0.3cm}
}

\date{MNRAS Submission}
\vspace{-0.7cm}

\pubyear{2021}

\begin{document}
\label{firstpage}
\pagerange{\pageref{firstpage}--\pageref{lastpage}}
\maketitle

\begin{abstract}

Well-characterised M-dwarfs are rare, particularly with respect to effective temperature. In this letter we re-analyse two benchmark M-dwarfs in eclipsing binaries from Kepler/K2: KIC 1571511AB and HD 24465AB. Both have temperatures reported to be hotter or colder by $\approx1000$ K in comparison with both models and the majority of the literature. By modelling the secondary eclipses with both the original data and new data from TESS we derive significantly different temperatures which are {\it not} outliers. Removing this discrepancy allows these M-dwarfs to be truly benchmarks. Our work also provides relief to stellar modellers. We encourage more measurements of M-dwarf effective temperatures with robust methods.

\end{abstract}




\vspace{0.3cm}

\section{Introduction}\label{sec:introduction}

There is a lack of precisely-characterised M-dwarfs in the literature. This inhibits our ability to constrain models of stellar structure for low mass stars. Exoplanet studies are also hampered, because our knowledge of the planets is limited by our knowledge of the host star. In the era of TESS and JWST, where M-dwarfs are popular targets, this is particularly problematic. In addition to poor statistics, there exist discrepancies between observations and theory. The most thoroughly studied is the so-called ``radius inflation'' problem, where M-dwarfs have been often observed with radii a few per cent higher at a given mass than expected by theoretical models \citep{Chabrier2000,Torres2014}. In this paper we tackle a different yet just as fundamental property: effective temperature. Like with the mass-radius relationship, we expect M-dwarfs to follow a mass-temperature relationship, with more massive stars being expectantly hotter. So far there is largely consistency between observations and theory, but any outliers must be rigorously studied.

Eclipsing binaries remain the most robust avenue for precise M-dwarf characterisation (e.g. \citealt{Triaud2017,vonBoetticher2019}). We can measure M-dwarf temperatures if we can observe the occultation of the M-dwarf by the companion star. In our study the M-dwarf is the smaller and cooler star in the binary, so its occultation is referred to as the secondary eclipse. An M-dwarf in the G + M eclipsing binary EBLM J0113+31 was found to have an effective temperature of $3922\pm42$ K \citep{MaqueoChew2014}, roughly 600 K hotter than expected for a $0.186M_\odot$ star. This irregularity was later shown to be erroneous by \citet{Swayne2020}, who calculated $T_{\rm eff,B}=3208\pm 40$ K, in line with expectations. The difference between the two studies was that \citet{Swayne2020} used TESS space-based photometry, whereas \citet{MaqueoChew2014} only had ground-based photometry. It was suggested that systematic errors in the J-band photometry created the error. A third analysis, \citet{Maxted2022}, added CHEOPS photometry and near-infrared SPIRou radial velocities. They derived a slightly higher temperature than \citet{Swayne2020} of $3375\pm40$ K. However, this is consistent with their heavier mass measurement of $0.197M_\odot$.

\begin{figure}
    \includegraphics[width=0.45\textwidth]{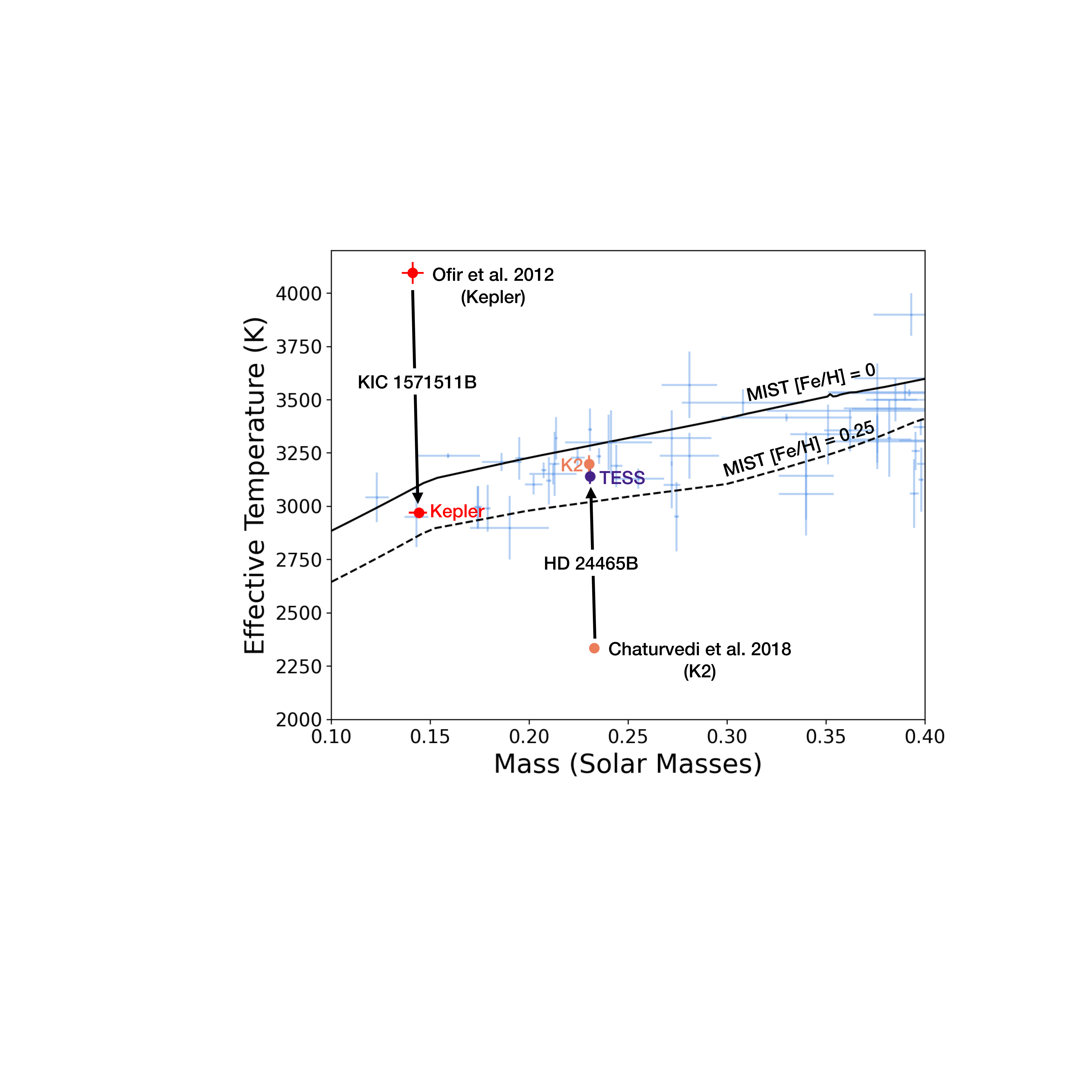}
    \caption{M-dwarfs from the literature (light blue) with precise mass and $T_{\rm eff}$ measurements. Two outliers with anomalously hot or cold temperatures: KIC 1571511B \citep{Ofir2012} and HD 24465B \citep{Chaturvedi2018} are highlighted, showing both the original published values and arrows pointing to our new values. The data are coloured according to the mission: Kepler (red), K2 (orange) and TESS (purple). The solid and dashed lines are theoretical model predictions from MIST \citep{mist}. Our temperatures are more in line with both theoretical models and the rest of the literature.}
    \label{fig:literature}
\end{figure}

In this paper we study two other benchmark M-dwarfs with outlier temperatures: KIC 1571511B \citep{Ofir2012} and HD 24465B \citep{Chaturvedi2018}. The outlier nature of these targets is demonstrated in Fig.~\ref{fig:literature}. KIC 1571511B has $M_{\rm B}=0.14136M_\odot$ and $T_{\rm eff,B}=4030 - 4150$, which is $\sim1000$ K {\it hotter} than expected from models and the bulk of the literature. Conversely, HD 24465B has $M_{\rm B}=0.233M_\odot$ and $T_{\rm eff,B}=2335.6\pm8.6$ K, which makes it $\sim 800$ K {\it colder} than expected. This temperature is also suspiciously precise compared with the rest of the literature. Both KIC 1571511AB and HD 24465AB were first analysed using space-based photometry (Kepler and K2, respectively). They both have clearly visible secondary eclipses. We both re-analyse the existing data and new TESS data, using methods applied in several existing studies \citep{Gill2019,Swayne2020,Swayne2021}. We demonstrate that, as was the case with EBLM J0113+31, the original published temperatures are  erroneous. We derive M-dwarf temperatures in line with theoretical models and the rest of the literature.

\begin{figure*}
    \includegraphics[width=0.99\textwidth]{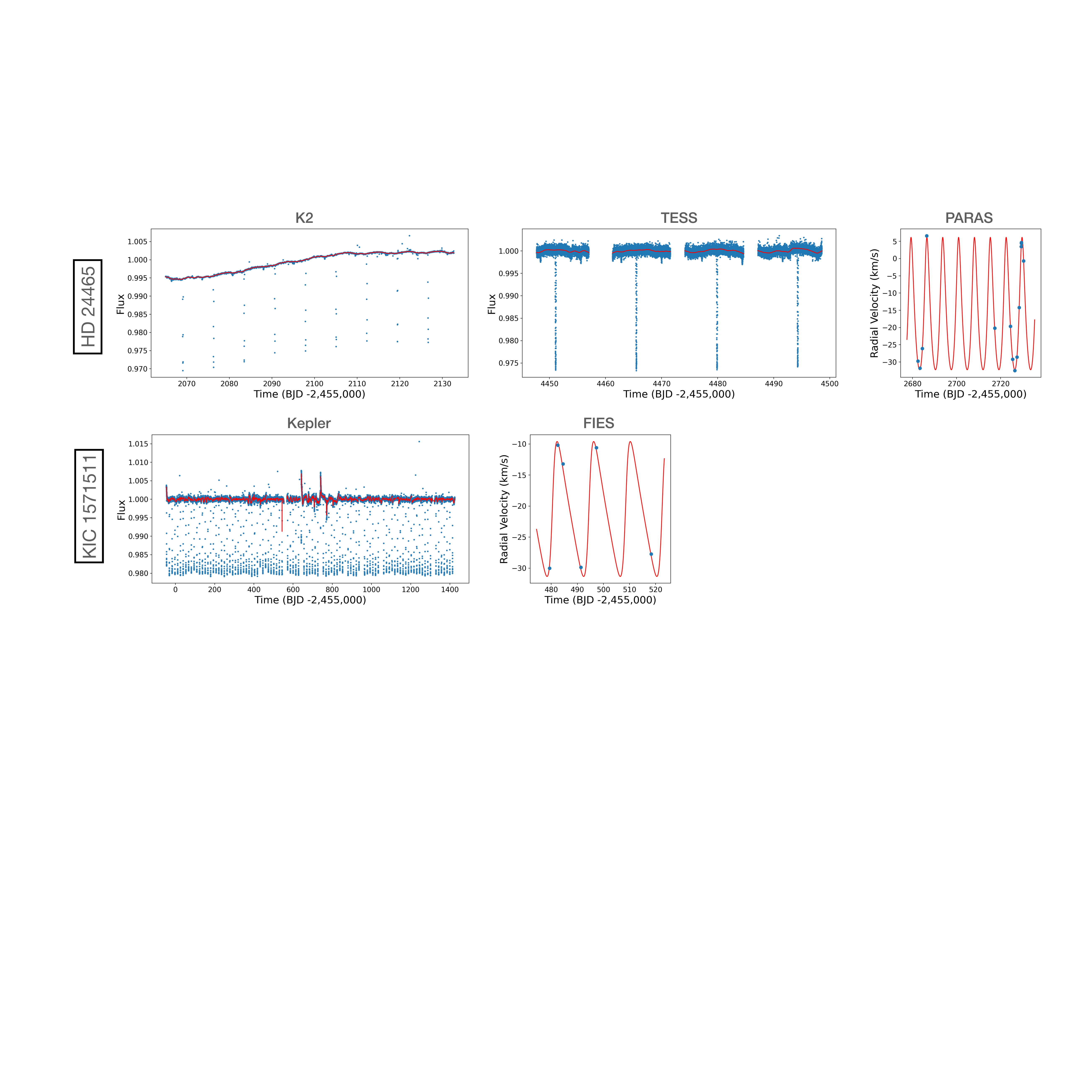}
    \caption{Photometry and radial velocity data for HD 24465AB (top row) and KIC 1571511AB (bottom row). K2 data are from the EVEREST pipeline \citet{Luger2016}. TESS and Kepler data are from the PDCSAP pipeline. KIC 1571511AB has TESS data but not of useable quality. The red line in the photometry is the fitted \textsc{Wotan} trend \citep{Hippke2019}. The radial velocity data are shown with the fitted Keplerian from \textsc{Exoplanet} \citep{foremanmackey2021}.}\label{fig:data}
\end{figure*}

\section{Targets and Observations}\label{sec:targets_observations}

Observational and stellar properties are cataloged in Table~\ref{tab:target_table}. The photometric and spectroscopic data are shown in Fig.~\ref{fig:data}.

\subsection{KIC 1571511AB}

This is a 14.0-day eclipsing binary containing $1.265M_\odot$ and $0.141M_\odot$ stars, discovered using data from the original Kepler mission \citet{Ofir2012}.  KIC 1571511B is considered a ``benchmark'' M-dwarf, which we define as having mass and radius errors less than 5\%. For KIC 1571511B: $\delta M_{\rm B}/M_{\rm B}=3.18\%$ and $\delta R_{\rm B}/R_{\rm B}=0.78\%$. The secondary star mass comes from six RV measurements from the FIbre-fed Echelle Spectrograph (FIES) on the Nordic Optical Telescope (NOT). \citet{Ofir2012} derive a temperature of $T_{\rm eff,B}=4030 - 4150$, which is roughly 1000 K hotter than expected. KIC 1571511AB has since been observed by the TESS space telescope. Unfortunately, the faintness of the target (Tmag = 12.95) means that we can see primary but not secondary eclipses, so we do not use these data.


\subsection{HD 24465AB}

\begin{figure*}
    \includegraphics[width=0.99\textwidth]{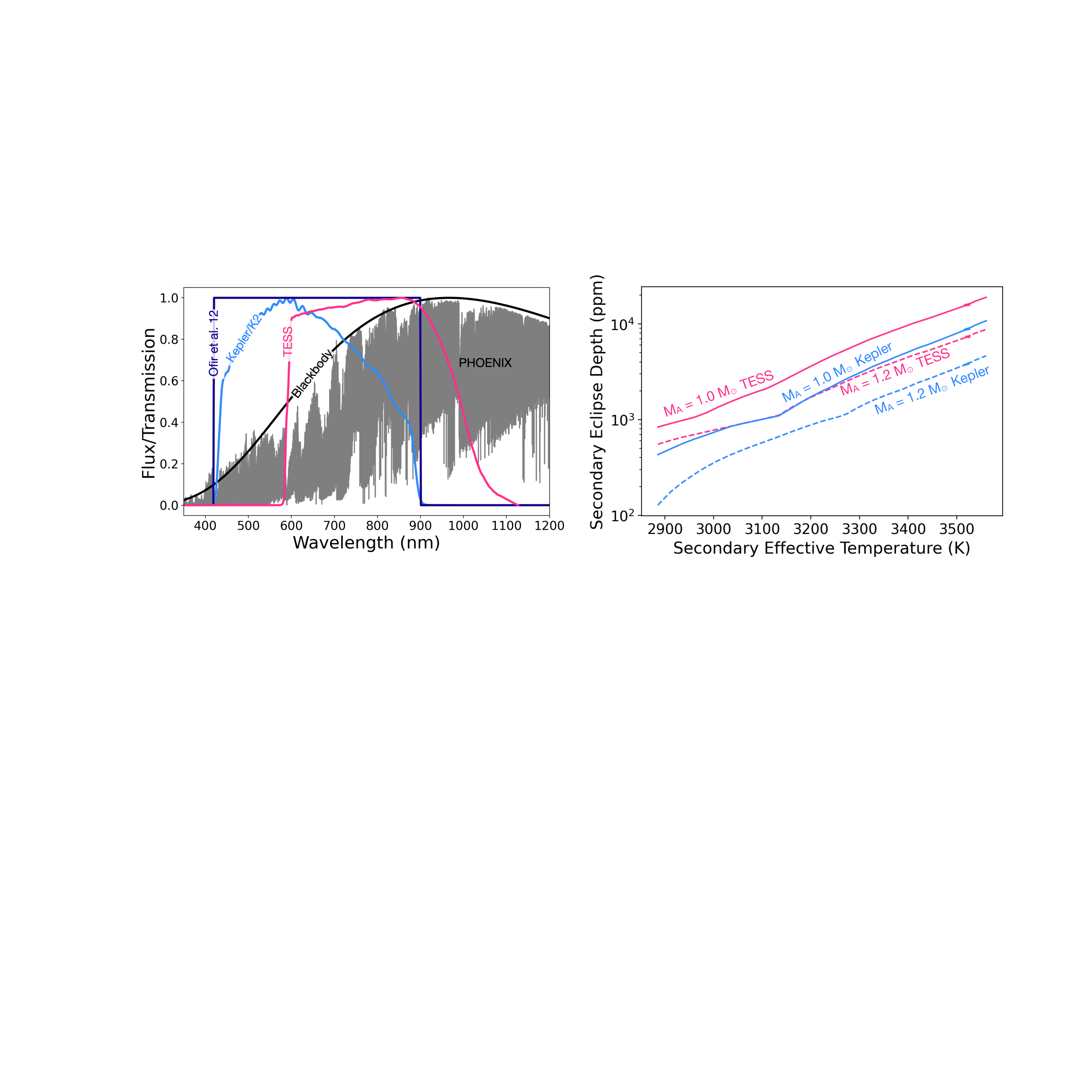}
    \caption{{\bf Left:} Transmission bandpasses for Kepler/K2 (blue) and TESS (pink), normalised to a maximum at 1. In navy is the \citet{Ofir2012} assumption of a  uniform Kepler transmission function. In grey is the \textsc{PHOENIX} model atmosphere for a 3000 K M-dwarf. In black is a blackbody curve, also for 3000 K. M-dwarfs have more flux at redder wavelengths, and hence we expect secondary eclipses to be deeper in TESS than in Kepler. We see this effect in HD 24465AB. {\bf Right:} predicted secondary eclipse depths for an M-dwarf as a function of $T_{\rm eff,B}$ around a $1.0M_\odot$ (solid lines) and $1.2M_\odot$ (dashed lines) G star, calculated using Eq.~\ref{eq:integral} for the Kepler (blue) and TESS (pink) bandpasses, assuming [Fe/H]$=0$.}\label{fig:bandpass}
\end{figure*}

This target comes from the \citet{Chaturvedi2018} study of four eclipsing binaries containing M-dwarfs. It is the only one which can be truly considered a benchmark M-dwarf, owing to precise K2 photometry. HD 24465AB is a 7.20-day binary consisting of $1.337M_\odot$ and $0.233M_\odot$ stars, where the M-dwarf is constrained to a precision of  $\delta M_{\rm B}/M_{\rm B}=0.86\%$ and $\delta R_{\rm B}/R_{\rm B}=0.4\%$, although we suspect that these errors do not properly account for the modelling uncertainties in the primary star's parameters (Duck et al. under rev.). The mass is derived from 14 radial velocities taken with the PARAS (PRL Advanced Radial-velocity Abu-sky Search) spectrograph on the 1.2-m telescope at Gurushikhar, Mount Abu, India. HD 24465AB was observed by TESS in sectors 42 and 43, both in short cadence (120s). Unlike for KIC 1571511AB, these data are sensitive to the secondary eclipse because this is a much brighter target (Tmag = 8.50). This provides an opportunity to measure the secondary eclipse depth in two different passbands, since TESS has a significantly redder sensitivity than Kepler (Fig.~\ref{fig:bandpass}). 

\renewcommand{\arraystretch}{1.3}

\begin{table}
\caption{Target information. Primary star parameters are taken from the original papers.}              
\label{tab:target_table}      
\centering                                    
\begin{tabular}{lll }  

\hline\hline                        
 
 Name
 & KIC 1571511AB
 & HD 24465AB  \\

\hline 
 
  TIC
 & 122680701
 & 242937935  \\
 
 $\alpha$
& $19^{\rm h}23^{'}59.256^{"}$ 
& $03^{\rm h}54^{'}03.371^{"}$ \\
& $290.9969^{\circ}$ 
& $58.5140^{\circ}$ \\

$\delta$ 
& $+37^{\circ}11' 57.18^{"}$
& $+15^{\circ}08' 30.19^{"}$\\
& $+37.1992^{\circ}$ 
& $+15.1417^{\circ}$ \\

Original Paper
& \citet{Ofir2012} 
& \citet{Chaturvedi2018} \\

$M_{\rm A}$ ($M_\odot$)
& $1.265^{+0.036}_{-0.030}$
& $1.337\pm0.008$\\

$R_{\rm A}$ ($R_\odot$)
& $1.343^{+0.012}_{-0.010}$
& $1.444\pm0.004$\\

$T_{\rm eff, A}$ (K)  
& $6195\pm50$ 
& $6250\pm100$ \\

${\rm [Fe/H]}$
& $0.37\pm0.08$
& $0.30\pm0.15$ \\
\hline

\end{tabular}
\end{table} 


\section{Methods}\label{sec:methods}

\begin{figure*}
    \includegraphics[width=0.85\textwidth]{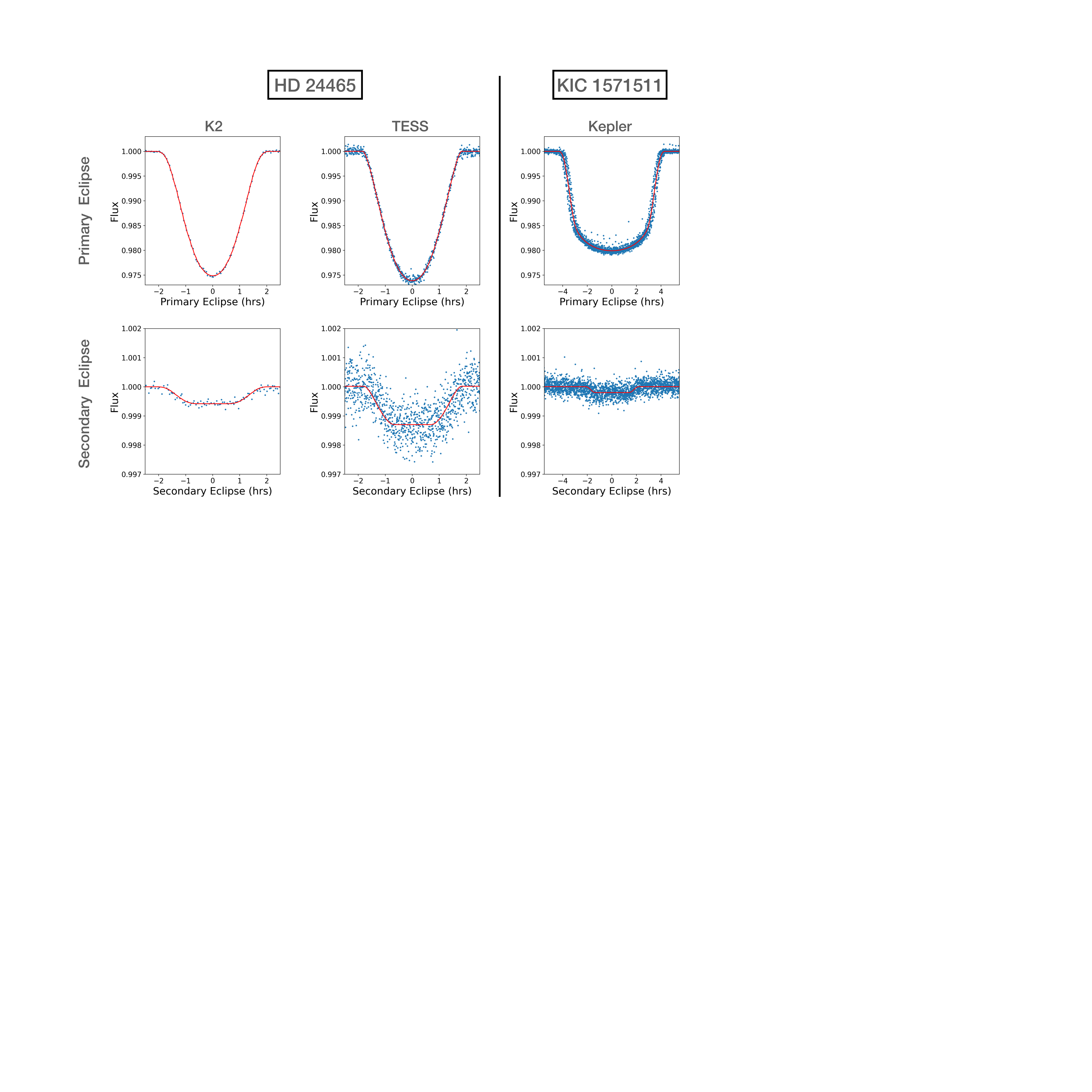}
    \caption{Exoplanet fits (red line) to the primary and secondary eclipse data (blue dots). To help comparison, in each row the vertical scale is the same across all of the targets. For HD 24465AB the secondary eclipse is significantly deeper in TESS because the M-dwarf emits more light at redder wavelengths and TESS has a redder bandpass than Kepler (Fig.~\ref{fig:bandpass}).}\label{fig:fits}
\end{figure*}

\subsection{Lightcurve Processing}\label{subsec:method_lightcurve}

For the Kepler, K2 and TESS data we use the \textsc{lightkurve} software package \cite{lightkurve2018} to download the data. For KIC 1571511AB we use the Kepler PDCSAP flux. For HD 24465AB we use the EVEREST flux \citep{Luger2016} for K2 and the PDCSAP flux for TESS. We flatten all three lightcurves using the \textsc{Wotan} detrending software \citep{Hippke2019}. We apply a Tukey's biweight filter with a 1 day window length, such that the eclipse depths are not affected. For HD 26645AB we manually removed the first few days of K2 data ($T-2,455,000 < 2065$). The original light curves and the fitted trends are shown in Fig~\ref{fig:data}.


\subsection{Exoplanet Fit}\label{subsec:methods_exoplanet}

We use the \textsc{exoplanet} software \citep{foremanmackey2021} to create joint photometry and radial-velocity fits. The fitted light curve, with primary and secondary eclipses, is calculated using \textsc{starry} \citep{Luger2019}, with quadratic limb darkening parameters calculated using \citet{Kipping2013}. After first estimating the maximum a posteriori parameters, we derive a posterior distribution and $1\sigma$ errorbars using PyMC3.

To convert from direct observables (e.g. the radial velocity semi-amplitude $K$ and the eclipse depths) to physical parameters ($M_{\rm B}$ and $R_{\rm B}$) we use the primary star mass and radius from the discovery papers (Table~\ref{tab:target_table}). These are implemented as a fixed value in all of the \textsc{Exoplanet} fits. The error in the primary star parameters is propagated to the errors in the secondary star mass and radius. The reason why we do not do a complete re-fit of all of the stellar and orbital parameters is that we want to fix as many parameters as possible. This will make it easier to identify the source of the surprisingly hot/cold temperatures previously published for the two targets.

For HD 22465 we do separate fits for the K2 and TESS data because the secondary eclipse depth will change in different passbands (Fig.~\ref{fig:bandpass}) and we seek two individual temperature measurements.


We note one issue with the radial velocity fits to HD 24465AB. We were unable to exactly replicate the fit of \citet{Chaturvedi2018} with their published data. In particular, our values for $K$ differ by $\approx 700$ m/s. In their Fig. 2 there are essentially no residuals to the RV fit, but in our best \textsc{exoplanet} fit we have residuals of 100's of m/s. We attempted an RV-only fit with the genetic algorithm \textsc{yorbit} \citep{Segransan2011}, but obtained the same fit as with \textsc{exoplanet}. Ultimately, our derived value for $M_{\rm B}$ is consistent with theirs, so for the purposes of this paper exploring the $T_{\rm eff}$ vs $M$ relationship our fit is sufficient. We had no such issues with the radial velocity fits of KIC 1571511AB. 

\subsection{M-dwarf Effective Temperature Derivation}\label{subsec:methods_temperature}

The secondary eclipse depth is related to the brightness ratio of the two stars by

\begin{align}
    \label{eq:sec_eclipse_depth}
    D_{\rm sec} &= k^2S + A_{\rm g} \left(\frac{R_{\rm B}}{a}\right)^2,
\end{align}
where $k$ is the radius ratio, $S$ is the surface brightness ratio and $A_{\rm g}$ is the geometric albedo \citep{Charbonneau2005,Canas2022}. In the first line the  $k^2S$ factor is the contribution from the intrinsic brightness of the M-dwarf. The $A_{\rm g}(R_{\rm B}/a)^2$ factor is light from the primary star reflected off the M-dwarf. Owing to the relatively wide separation of the binaries and a typical albedo of $A_{\rm g}=0.1$ \citep{Marley1999,Canas2022}, the reflection effect can be considered negligible. For example, for KIC 1571511 the reflection effect is $\approx4$ ppm, relative to a $\approx200$ ppm secondary eclipse. 

To derive the secondary star's effective temperature we first calculate $S$ from Eq.~\ref{eq:sec_eclipse_depth} and then solve for $T_{\rm eff,B}$ in

\begin{align}
    \label{eq:integral}
    S= \frac{\int \tau(\lambda)F_{\rm B,\nu}(\lambda,T_{\rm eff,B},\log{g_{\rm B}})\lambda d\lambda}{\int \tau(\lambda)F_{\rm A,\nu}(\lambda,T_{\rm eff,A},\log{g_{\rm A}})\lambda d\lambda},
\end{align}
where $\tau$ is the instrumental transmission function for Kepler/K2/TESS\footnote{Kepler and K2 are different missions but the same telescope, and hence the same transmission function. Both Kepler and TESS transmission functions can be downloaded here: \url{http://svo2.cab.inta-csic.es/svo/theory/fps3/index.php?}} and $F$ is the flux of each star as a function of wavelength $\lambda$,  effective temperature and surface gravity.
The factor of $\lambda$ inside each integral is the same correction as made in Duck et al. (under rev.), based on \citet{Bessell2012}. This is because the transmission functions are setup for the photon-counting instrumental CCDs, and so we add a factor of $\lambda/(hc)$ to calculate the instrumental flux rather than the photon count. The constants $hc$ are added to both integrals and hence cancel.
We calculate $F$ using an interpolated grid of PHOENIX stellar spectra models \citep{husser2013}.  In Eq.~\ref{eq:integral} we convolve these theoretical spectra with the instrument's bandpass to predict the star's observed brightness. This is done for both stars. For the primary star we take the published value of $T_{\rm eff,A}$ and $\log{g_{\rm A}}$. For the secondary star we take our fitted value of $\log{g_{\rm B}}$, use the literature value for [Fe/H] and test a grid of $T_{\rm eff,B}$ between 2500 and 4000 K. We then solve Eq.~\ref{eq:integral} for $T_{\rm eff}$. The error bar on $T_{\rm eff,B}$ comes from applying Eq.~\ref{eq:integral} with the $1\sigma$ errors on $S$, [Fe/H], $T_{\rm eff,A}$, $\log{g_{\rm A}}$ and $\log{g_{\rm B}}$.

In Fig.~\ref{fig:bandpass} (right) we demonstrate how the secondary eclipse depth changes as a function of $T_{\rm eff,B}$, for different bandpasses (Kepler and TESS) and different host star masses ($1.0M_\odot$ and $1.2M_\odot$). We see that detecting secondary eclipses for late M-dwarfs (<3000 K) becomes very challenging.





\section{Results and Discussion}\label{section:results_discussion}

 We derive effective temperatures that are significantly different to the \citet{Ofir2012,Chaturvedi2018} results, but in line with expectations from both models and the rest of the literature. This difference is highlighted in Fig.~\ref{fig:literature}. All of the results provided in Table~\ref{tab:params}. In Fig.~\ref{fig:fits} we show zoomed fits to the primary and secondary eclipses for both targets.
 
 For HD 24465AB our TESS and K2 temperatures are slightly discrepant at the $\approx2\sigma$ level. This may be an artefact of our handling of dilution or light curve detrending. Our fractional uncertainty on $D_{\rm sec}$ for HD24465AB is 0.4\% for K2, compared with 1.7\% for TESS. The higher precision of K2 more than compensates for the deeper secondary eclipse in TESS's redder bandpass (Fig.~\ref{fig:bandpass}).
Our differences between $T_{\rm eff,B}$ in K2 and TESS are very small relative to the difference with the value from \citet{Chaturvedi2018}. Our fitted parameters for the M-dwarf mass and radius, as well as the binary orbital parameters, largely match the discovery papers. This suggests consistency with the fitting of the radial velocities and the primary eclipse, at least. 

Why was the \citet{Ofir2012} temperature for KIC 1571511B roughly $1000$ K {\it too hot}, and the \citet{Chaturvedi2018} result  for HD 24465B roughly $800$ K {\it too cold}?


\begin{table*}
\caption{Fitted parameters from both this paper and the literature.} 
\begin{tabular}{l|ll|lll}
\hline
Target           & \multicolumn{2}{c|}{KIC 1571511AB}                                                    & \multicolumn{3}{c}{HD 24465AB}                                                                  \\\hline\hline
Author           & This Paper                               & \citet{Ofir2012}                        & This Paper & This Paper                                    &   \citet{Chaturvedi2018}  \\

Instrument       & Kepler                                   & Kepler                                  & TESS                            & K2                              & K2                        \\
\hline 
$M_{\rm B}$ ($M_\odot$)    & $0.1441\pm0.0025$             & $0.1414^{+0.0051}_{-0.0042}$   & $0.23077\pm0.00092$  & $0.23020\pm0.00092$  & $0.233\pm0.002$  \\
$R_{\rm B}$ ($R_\odot$)     & $0.1770\pm0.0014$             & $0.1783^{+0.0013}_{-0.0016}$   & $0.2475\pm0.00069$      & $0.24235\pm0.00069$    & $0.244\pm0.001$  \\
$P$ (days)              & $14.022640\pm0.000000052$ & $14.02248^{+0.000023}_{-0.000021}$  & $7.196365\pm0.000002$       & $7.19644\pm0.000002$        & $7.19635\pm0.00002$   \\
$e$              & $0.3661\pm0.0015$                        & $0.3269 \pm 0.0027$                     & $0.20792\pm0.00016$             & $0.20948\pm0.00010$             & $0.208\pm0.002$           \\
$K$ (km/s)              & $10.515\pm0.037$                                      & $10.521\pm0.024$                    & $19.227\pm0.006$                             & $19.307\pm0.006$                             & $18.629\pm0.053$      \\
$b_{\rm pri}$              & $0.3737\pm0.0052$                        & $0.383^{+0.040}_{-0.049}$                                     & $0.8584\pm0.0042$               & $0.84161\pm0.00067$             & *$0.83926$                       \\
$k$              & $0.13180\pm0.00014$                      & $0.13277^{+0.00038}_{-0.00046}$         & $0.1714\pm0.0015$               & $0.16783\pm0.00015$             & *$0.169$                       \\
$D_{\rm sec}$ (normalised flux)              & $0.0002043\pm0.0000059$                      & $0.000275\pm0.000019$                                     & $0.001340\pm0.000023$             & $0.0005812\pm0.0000023$           & 0.000018                       \\
$D_{\rm sec}$ (ppm)              & $204.3\pm5.9$                      & $275\pm19$                                     & $1340\pm23$             & $581.2\pm2.3$           & 18                       \\
$S$              & $0.01176\pm0.00034$                      & *0.01560                                     & $0.04476\pm0.00098$             & $0.020633\pm0.000073$           & *0.00063                       \\
$T_{\rm eff, B}$ Observed (K) & $2970\pm17$                                      & $4030 - 4150$                          & $3142\pm36$                             & $3200\pm38$                             & $2335.60\pm8.56$      \\ 
\hline
$T_{\rm eff, B}$ MIST Model (K) & \multicolumn{2}{c|}{2863} & \multicolumn{3}{c}{3020} \\
\hline
\end{tabular}
\flushleft\footnotesize{Parameter descriptions in order: $M_{\rm B}$ - M-dwarf mass; $R_{\rm B}$ - M-dwarf radius; $P$ - binary period; $e$ - binary eccentricity; $K$ - radial velocity semi-amplitude; $b_{\rm pri}$ - primary eclipse impact parameter; $k=R_{\rm B}/R_{\rm A}$ - radius ratio; $D_{\rm sec}$ - secondary eclipse depth in normalised flux units; $S$ surface brightness ratio; $T_{\rm eff,B}$  Observed - M-dwarf effective temperature; $T_{\rm eff,B}$ MIST Model - theoretically predicted temperature from MIST stellar models \citep{mist}. Parameters noted with *  were not explicitly provided in the earlier papers and are instead calculated by us. We suspect that $D_{\rm sec}=$ 18 ppm ``observed'' by \citet{Chaturvedi2018} was actually a predicted value from their \textsc{PHOENIX} fit of the primary eclipse.}
\label{tab:params}
\end{table*}



\subsection{KIC 1571511B}

\citet{Ofir2012} derive an M-dwarf temperature of 4030 - 4150 K, which is more than 1000 K hotter than our value of $2970\pm 17$ K. There are three differences between our studies. First, their secondary eclipse depth ($D_{\rm sec}=0.000275\pm0.000019$) is roughly $3\sigma$ deeper than ours ($D_{\rm sec}=0.0002043\pm0.0000059$), despite both values being derived from Kepler data. It is possible that different light curve processing led to this discrepancy. We re-do our analysis with the \citet{Ofir2012} $D_{\rm sec}$ and derive only a slightly hotter temperature of $3075\pm 18K$, which would still be within theoretical expectations.

A second difference is that \citet{Ofir2012} derive $T_{\rm eff,B}$ assuming blackbodies, as opposed to our \textsc{PHOENIX} model spectra (comparison in Fig.~\ref{fig:bandpass}). We re-do our analysis with this assumption and actually obtain a colder temperature of $2695\pm 15$ K. This would be an outlier, but in the opposite direction of the \citet{Ofir2012} result. The third difference is that \citet{Ofir2012} assume a uniform Kepler passband between 420 and 900 nm. By applying this assumption we obtain $2884\pm15$. Again, this simplying assumption actually acts to make the M-dwarf {\it cooler}. This can be seen in Fig.~\ref{fig:bandpass} where the assumption of a uniform bandpass would imply a higher Kepler sensitivity at redder wavelengths. Overall, we cannot explain why the \citet{Ofir2012} result is so hot.

\subsection{HD 26645}

\citet{Chaturvedi2018} derive $T_{\rm eff,B}=2335.60\pm8.56$, which is 865 K cooler than our value from K2. We suspect that this value does not come from fitting the secondary eclipse, since \citet{Chaturvedi2018} note ``The secondary eclipse depth for all the sources were either undetectable or small'', yet they provide measurements for $T_{\rm eff,B}$ in all four cases. Figure~\ref{fig:fits} show that the secondary eclipse is in fact clear in our K2 data. However, \citet{Chaturvedi2018} do not appear to be using the EVEREST pipeline, and hence the secondary eclipse may have been hidden to them behind telescope systematics.

\citet{Chaturvedi2018} use the \textsc{PHOEBE} \citep{Prsa2011Phoebe} package to model the photometry and radial velocities, and state that $T_{\rm eff,B}$ is ``kept free for fitting''. However, since the fit was seemingly only of the {\it primary} eclipse and the radial velocities, there is essentially no information in the light curve concerning $T_{\rm eff,B}$. \textsc{PHOEBE} is arguably the most detailed software available for fitting eclipsing binaries, and we doubt it would produce such an outlier $T_{\rm eff,B}$ if both eclipses were being fit.

Finally, \citet{Chaturvedi2018} do note for HD 24465AB a secondary eclipse depth of 0.000018 in normalised flux units. This value is 32 times smaller than ours. We suspect this is not an observed depth but a predicted depth based on the $2335$ K effective temperature.

\section{Conclusion}\label{section:conclusion}

We have studied two benchmark M-dwarfs: KIC 1571511B \citep{Ofir2012} and HD 24465B \citep{Chaturvedi2018}. The former had a reported temperature about 1000 K hotter than expected. The latter was about 800 K colder than expected. Such discoveries would have posed significant challenges to stellar models. We re-analyse the original Kepler/K2 data to derive the M-dwarf effective temperature based on the secondary eclipse depth. Our results differ significantly from the original studies, and instead match the temperatures expected from both models and the majority of  other literature M-dwarfs. With these new precise and reliable M-dwarf temperatures, these two targets can be truly considered  benchmarks.

\section*{Data Availability Statement}

All radial velocities and light curves will be made available online.

\section*{Acknowledgements}

Support for DVM was provided by NASA through the NASA Hubble Fellowship grant HF2-51464 awarded by the Space Telescope Science Institute, which is operated by the Association of Universities for Research in Astronomy, Inc., for NASA, under contract NAS5-26555. This research is also supported work funded from the European Research Council (ERC) the European Union’s Horizon 2020 research and innovation programme (grant agreement n◦803193/BEBOP).  Partial support for AD, RRM, and BSG
was provided by the Thomas Jefferson Chair Endowment for Discovery and Space Exploration. 
MIS acknowledges support from STFC grant number ST/T506175/1.

\bibliographystyle{mnras}
\bibliography{outlier_temps} 


\bsp	
\label{lastpage}
\end{document}